\documentstyle [12pt] {article}

\catcode`@=12
\begin{document}
\thispagestyle{empty}
\begin{flushright}
UCRHEP-T124\\
April 1994\\
\end{flushright}
\vspace{1.0in}
\begin{center}
{\large \bf Left-Right Symmetry and Supersymmetric Unification\\}
\vspace{1.5in}
{\bf Ernest Ma\\}
\vspace{0.1in}
{\sl Department of Physics\\
University of California\\
Riverside, California 92521\\}
\vspace{1.5in}
\end{center}
\begin{abstract}\
The existence of an $\rm SU(3) \times SU(2)_L \times SU(2)_R \times U(1)$
gauge symmetry with $g_{\rm L} = g_{\rm R}$ at the TeV energy scale is shown
to be consistent with supersymmetric SO(10) grand unification at around
$10^{16}$ GeV if certain new particles are assumed.  The additional imposition
of a discrete $\rm Z_2$ symmetry leads to a generalized definition of R parity
as well as highly suppressed Majorana neutrino masses.  Another model
based on $\rm SO(10) \times SO(10)$ is also discussed.
\end{abstract}

\newpage
\baselineskip 24pt

Possible unification of the three gauge couplings of the standard
$\rm SU(3)_C \times SU(2)_L \times U(1)_Y$ model of quark and lepton
interactions in its minimal supersymmetric extension has revived
widespread interest in this topic.\cite{1}  The success of such a
scenario seems to imply that no new physics beyond the minimal
supersymmetric standard model (MSSM) would be observed below an energy
scale of a few TeV.  In particular, the left-right gauge extension
appears to be excluded, except for one very special case,\cite{2} but
even then, the condition $g_{\rm L} = g_{\rm R}$ cannot be maintained.\cite{3}
If $g_{\rm L} = g_{\rm R}$ is desired, the scale $\rm M_R$ of
$\rm SU(2)_R$ symmetry breaking obtained in previous studies\cite{4}
is typically of the order $10^{10}$ GeV.  The purpose of this paper is
to show in an explicit first example that it is actually possible
to have supersymmetric SO(10) grand unification at around $10^{16}$ GeV
as well as an $\rm SU(3) \times SU(2)_L \times SU(2)_R \times U(1)$ gauge
symmetry with $g_{\rm L} = g_{\rm R}$ at the TeV energy scale.
New particles will be assumed together with a discrete Z$_2$
symmetry, resulting in a generalized definition of R parity together with
highly suppressed Majorana neutrino masses.  Another model based on
SO(10) $\times$ SO(10) will also be discussed.

Consider a supersymmetric SO(10) model which breaks down at the
unification scale $\rm M_U$ to a supersymmetric $\rm SU(3) \times
SU(2)_L \times SU(2)_R \times U(1)$ model with left-right exchange
symmetry, which then breaks down at $\rm M_R$ to the standard
(nonsupersymmetric) $\rm SU(3) \times SU(2)_L \times U(1)_Y$ model.
The scale of supersymmetry breaking is assumed to coincide with $\rm M_R$.
Before going into the details of the evolution of the various gauge
couplings, note that with a left-right model, there should be two
scalar bidoublets in order that realistic quark and lepton mass
matrices be obtained.  Hence there are four $\rm SU(2)_L$ doublets
which will change the evolution of the $\rm SU(2)_L$ coupling $g_{\rm L}$
but not that of the SU(3) coupling $g_{\rm S}$, leading to the loss of
unification of the gauge couplings.  In the MSSM, this corresponds to
the well-known fact that unification occurs with two, but not four,
Higgs doublets.  Note also that this problem exists whether or not
$g_{\rm L} = g_{\rm R}$ unless $\rm M_R = M_U$.  Consequently, new
particles are necessary to offset the effect of the extra bidoublet if
$\rm SU(2)_R$ is to be a gauge symmetry already above a few TeV.

Consider now the evolution of the gauge couplings to one-loop order.
Generically,
\begin{equation}
\alpha_i^{-1}(M_1) = \alpha_i^{-1}(M_2) - {b_i \over {2\pi}} \ln {M_1
\over M_2},
\end{equation}
where $\alpha_i \equiv g_i^2/4\pi$ and $b_i$ are constants determined
by the particle content contributing to $\alpha_i$.  The initial conditions
are set at $\rm M_Z = 91.187 \pm 0.007$ GeV\cite{5} by the experimental
values $\alpha^{-1} = 127.9 \pm 0.1$,\cite{6} $\sin^2 \theta_{\rm W} =
0.2321 \pm 0.0006$,\cite{7} and $\alpha_{\rm S} = 0.120 \pm 0.006 \pm
0.002$.\cite{8}  Hence
\begin{equation}
\rm \alpha_S^{-1} (M_Z) = 8.33 \begin{array} {c} + 0.60 \\ - 0.52
\end{array},
\end{equation}
\begin{equation}
\rm \alpha_L^{-1} (M_Z) = 29.69 \pm 0.10,
\end{equation}
and
\begin{equation}
\rm \alpha_Y^{-1} (M_Z) = 98.21 \pm 0.15.
\end{equation}
At $\rm M_R$, the matching conditions of the gauge couplings are
\begin{equation}
\rm \alpha_L^{-1} (M_R) = \alpha_R^{-1} (M_R),
\end{equation}
and
\begin{equation}
\rm \alpha_Y^{-1} (M_R) = \alpha_R^{-1} (M_R) + \alpha_X^{-1} (M_R),
\end{equation}
where $\rm \alpha_X$ refers to the U(1) gauge coupling of the left-right
symmetry.  Above $\rm M_R$, $\rm \alpha_L$ and $\rm \alpha_R$ will evolve
together identically.

The particle content of this model is as follows.  There are three copies
of the {\bf 16} representation of SO(10) consisting of the usual quarks
and leptons.  Their transformations under $\rm SU(3) \times SU(2)_L \times
SU(2)_R \times U(1)$ are given by
\begin{equation}
Q \sim (3,2,1,1/6), ~~~~~ Q^c \sim (\overline {3}, 1,2,-1/6),
\end{equation}
and
\begin{equation}
L \sim (1,2,1,-1/2), ~~~~~ L^c \sim (1,1,2,1/2).
\end{equation}
There are two bidoublets
\begin{equation}
\Phi_{12} \sim (1,2,2,0),
\end{equation}
which are necessary for realistic quark and lepton mass matrices as
already mentioned, and one set of $\rm SU(2)_L$ and $\rm SU(2)_R$
doublets and their charge-conjugate partners
\begin{equation}
\rm \Phi_L \sim (1,2,1,-1/2), ~~~~~ \Phi_R \sim (1,1,2,1/2),
\end{equation}
and
\begin{equation}
\Phi_{\rm L}^c \sim (1,2,1,1/2), ~~~~~ \Phi_{\rm R}^c \sim (1,1,2,-1/2),
\end{equation}
so that $\rm SU(2)_R$ may be broken independently of $\rm SU(2)_L$.
Added to this minimal collection are two copies each of
\begin{equation}
D \sim (3,1,1,-1/3), ~~~~~ D^c \sim (\overline {3}, 1,1,1/3),
\end{equation}
and
\begin{equation}
E \sim (1,1,1,-1), ~~~~~ E^c \sim (1,1,1,1),
\end{equation}
as well as three copies of
\begin{equation}
N \sim (1,1,1,0).
\end{equation}
As shown below, this choice will allow $\rm M_R$ to be a few TeV with
$\rm M_U$ of the order $10^{16}$ GeV.

In the one-loop approximation, below $\rm M_R$,
\begin{eqnarray}
b_{\rm S} &=& -11 + {4 \over 3} (3) = -7,\\
b_{\rm L} &=& -{22 \over 3} + {4 \over 3} (3) + {1 \over 6} (2) = -3,\\
b_{\rm Y} &=& {20 \over 9} (3) + {1 \over 6} (2) = 7,
\end{eqnarray}
whereas above $\rm M_R$,
\begin{eqnarray}
b_{\rm S} &=& -9 + 2 (3) + n_D = -1,\\
b_{\rm LR} &=& -6 + 2 (3) + n_{22} + n_H = 3,\\
{3 \over 2} b_{\rm X} &=& 2(3) + 3n_H + n_D + 3n_E = 17,
\end{eqnarray}
where $n_{22} = 2$, $n_H =1$, $n_D = 2$, $n_E = 2$, and the factor 3/2
for $b_{\rm X}$ comes from the normalization of
the $\rm U(1)_X$ coupling within SO(10).  Assuming that $\rm \alpha_S^{-1}
(M_U) = \alpha_{LR}^{-1} (M_U) = (3/2) \alpha_X^{-1} (M_U)$, Eq. (1) can be
solved for $\rm M_R$ and $\rm M_U$, {\it i.e.}
\begin{equation}
\ln \rm {M_R \over M_Z} = {\pi \over 4} \left[ 3 \alpha^{-1} (M_Z)
\{1-5 sin^2 \theta_W (M_Z)\} + 7 \alpha_S^{-1} (M_Z) \right] < 1.66,
\end{equation}
and
\begin{equation}
\ln \rm {M_U \over M_Z} = {\pi \over 2} \left[ \alpha^{-1} (M_Z)
sin^2 \theta_W (M_Z) - \alpha_S^{-1} (M_Z) \right] > 32.45.
\end{equation}
Hence $\rm M_R < 480$ GeV and $\rm M_U > 1.1 \times 10^{16}$ GeV.
The upper bound of $\rm M_U$ is $1.9 \times 10^{16}$ GeV, corresponding
to $\rm M_R = M_Z$.

The allowed parameter space opens up more in two loops.  Using\cite{9}
\begin{equation}
b_{ij} = \left( \begin{array}{c@{\quad}c@{\quad}c} -26 & {9 \over 2} &
{11 \over 10} \\ 12 & 8 & {6 \over 5} \\ {44 \over 5} & {18 \over 5} &
{104 \over 25} \end{array} \right)
\end{equation}
for $\rm \alpha_S^{-1}$, $\rm \alpha_L^{-1}$, and $\rm (3/5) \alpha_Y^{-1}$
below $\rm M_R$, and
\begin{equation}
b_{ij} = \left( \begin{array}{c@{\quad}c@{\quad}c} {110 \over 3} & 9 &
{7 \over 3} \\ 24 & 45 & {9 \over 2} \\ {56 \over 3} & {27 \over 2} &
{293 \over 6} \end{array} \right)
\end{equation}
for $\rm \alpha_S^{-1}$, $\rm \alpha_{LR}^{-1}$, and $\rm (3/2) \alpha_X^{-1}$
above $\rm M_R$, and solving the equations
\begin{equation}
\mu {{\partial \alpha_i (\mu)} \over {\partial \mu}} = {1 \over {2\pi}}
\left( b_i + {b_{ij} \over {4\pi}} \alpha_j (\mu) \right) \alpha_i^2 (\mu)
\end{equation}
numerically with the proper boundary conditions at $\rm M_U$:
\begin{equation}
\rm \alpha_U^{-1} - {2 \over {3\pi}} = \alpha_S^{-1} - {1 \over {4\pi}} =
\alpha_{LR}^{-1} - {1 \over {6\pi}} = {3 \over 2} \alpha_X^{-1},
\end{equation}
it is found that
\begin{equation}
\rm 6.3 \times 10^{15}~GeV < M_U < 2.3 \times 10^{16}~GeV
\end{equation}
with
\begin{equation}
\rm 6.7~TeV > M_R > M_Z.
\end{equation}
As an example, Fig. 1 shows the case where $\rm M_R$ is chosen
arbitrarily to be 1 TeV with the central values
$\rm \alpha^{-1} (M_Z) = 127.9$ and $\rm sin^2 \theta_W (M_Z) = 0.2321$
as inputs, from which $\rm M_U = 1.0 \times 10^{16}$ GeV and
$\rm \alpha_S (M_Z) = 0.115$ are obtained.  Note that $\rm M_R$
depends very sensitively on $\rm \alpha_S$.  If $\rm \alpha_S (M_Z) =
0.120$ is used in the above, $\rm M_R$ drops down by an order of
magnitude to 120 GeV, whereas $\rm M_U$ increases only about twofold to
$1.8 \times 10^{16}$ GeV.

The matter superfields of this model are assumed to be distinguished
by a discrete $\rm Z_2$ symmetry where $\rm \Phi_R$, $\Phi_{\rm R}^c$,
and $\Phi_{12}$ are even, and all the others are odd.  This is merely
a generalization of the usual procedure in the MSSM where the two Higgs
superfields are chosen to be even, and the quark and lepton superfields
odd.  As a result, a generalized conserved R parity also exists in this
model.  Note that the terms $Q Q D$, $Q^c Q^c D^c$, $L Q D^c$, and
$L^c Q^c D$ are all forbidden.

The spontaneous breaking of the $\rm SU(2)_R \times U(1)_X$
gauge symmetry down to $\rm U(1)_Y$ is accomplished by the nonzero
vacuum expectation values of the neutral scalar components of $\rm \Phi_R$
and $\Phi_{\rm R}^c$.  Similarly, $\rm SU(2)_L \times U(1)_Y$ breaks
down to $\rm U(1)_Q$ via $\Phi_{12}$.  The usual quarks and leptons
obtain their masses through the Yukawa terms $Q Q^c \Phi_{12}$ and
$L L^c \Phi_{12}$.  The exotic quarks and leptons have gauge-invariant
mass terms $D D^c$ and $E E^c$.  Mixing between the two sectors occurs
through the terms $D Q^c \Phi_{\rm R}$ and $E L^c \Phi_{\rm R}$.  This
means that whereas $d^c - D^c$ and $e^c - E^c$ mixing may be substantial,
$d - D$ and $e - E$ mixing are guaranteed to be small.\cite{10}
Although $\rm \Phi_L$ and $L$ transform identically and neither have
any vacuum expectation value,
the gauge-invariant mass term $\Phi_{\rm L} \Phi_{\rm L}^c$ can be used
to define $\rm \Phi_L$, after which there is of course still an allowed
$\Phi_{\rm L} L^c \Phi_{12}$ term, but the $L - \Phi_{\rm L}$ mixing
will be highly suppressed.\cite{10}

Consider now the neutrino sector.  Ignoring the small mixing with the
neutral spinor components of $\rm \Phi_L$ and $\Phi_{\rm L}^c$, the
mass matrix spanning $\nu$, $\nu^c$, and $N$ is of the form
\begin{equation}
{\cal M} = \left( \begin{array}{c@{\quad}c@{\quad}c} 0 & m_{\rm D} & 0 \\
m_{\rm D} & 0 & m_{\rm R} \\ 0 & m_{\rm R} & m_N \end{array} \right),
\end{equation}
where $m_{\rm D}$ comes from the $L L^c \Phi_{12}$ term, $m_{\rm R}$
from the $N L^c \Phi_{\rm R}^c$ term, and $m_N$ is an allowed
gauge-invariant Majorana mass term for $N$.  Note that if $m_N = 0$,
then additive lepton number is conserved and the matrix $\cal M$ has
a zero eigenvalue which is the physical mass of a linear combination
of $\nu$ and $N$, while its orthogonal combination pairs up with
$\nu^c$ to form a heavy Dirac fermion of mass $\sqrt {m_{\rm R}^2 +
m_{\rm D}^2}$.  If $m_N$ is not zero but nevertheless small, $\nu$ will
pick up a very small Majorana mass given by
\begin{equation}
m_\nu \simeq {{m_N m_{\rm D}^2} \over m_{\rm R}^2}.
\end{equation}
Hence $m_\nu$ is not only suppressed by the usual seesaw mechanism,
but also by the small ratio $m_N/m_{\rm R}$.\cite{11}  This allows
$m_\nu$ to be very small even though $\rm M_R$ is only a few TeV.
For example, if $m_{\rm D} = 1$ MeV, $\rm M_R = 1$ TeV, and $m_N = 1$
GeV, then $m_\nu \simeq 10^{-3}$ eV.  Note that if $\cal M$ has a
$\nu N$ mass term $m_{\rm L}$, then $m_\nu$ has an additional
contribution $\simeq -2m_{\rm D} m_{\rm L}/m_{\rm R}$ which is not
suppressed by $m_N/m_{\rm R}$.  However, $m_{\rm L}$ is absent
because the $N L \Phi_{\rm L}^c$ term is forbidden by the assumed
discrete $\rm Z_2$ symmetry.  On the other hand, left-right exchange
symmetry is now broken in the Yukawa sector and the equality $g_{\rm L}
= g_{\rm R}$ is violated slightly due to the Yukawa contributions to
the renormalization-group equations of the gauge couplings, which first
appear in two loops.

Consider next an $\rm SO(10) \times SO(10)$ model which also breaks down
at $\rm M_U$ to $\rm SU(3) \times SU(2)_L \times SU(2)_R \times U(1)_X$
but the $\rm U(1)_X$ is now a linear combination of the usual U(1)
in the first SO(10) and an U(1) remnant of the second SO(10).  All
low-energy matter supermultiplets are assumed to be only those of the first
SO(10) as well as all gauge particles except for the photon (and photino)
which spans both SO(10)'s.  Specifically, the electric
charge is given by $Q_1 + Q_2$, where $Q_1$ and $Q_2$ are embedded in the
two SO(10)'s in exactly the same way.  The normalization
factor for $\rm \alpha_X^{-1}$ is then 3/10 instead of 3/2, because
$(3/10)^{-1} = (3/2)^{-1} + (3/8)^{-1}$, assuming of course that the
two SO(10) gauge couplings are equal at $\rm M_U$.

The particle content of this model is now assumed to consist of three
copies of the {\bf 16} representation, one copy of the {\bf 16*}
representation, two bidoublets, three $N$'s and one set of $D$ and $D^c$.
Below $\rm M_R$, the $b_i$'s are given by $b_{\rm S} = -11 + (4/3)(4)
= -17/3$, $b_{\rm L} = -22/3 + (4/3)(4) + (1/6)(2) = -5/3$, and
$b_{\rm Y} = (20/9)(4) + (1/6)(2) = 83/9$.  Above $\rm M_R$,
$b_{\rm S} = -9 + 2(4) + 1 = 0$, $b_{\rm LR} = -6 + 2(4) + 2 = 4$, and
$(3/10)b_{\rm X} = (2/5)(4) + 1/5 = 9/5$.  Solving Eq. (1) for
$\rm M_R$ and $\rm M_U$, it is easily seen that $\rm M_U$ is
again given by Eq. (22) whereas
\begin{equation}
\rm ln {M_R \over M_Z} = {{3\pi} \over 37} \left[ {3 \over 2} \alpha^{-1}
(M_Z) \{1 - {7 \over 2} sin^2 \theta_W (M_Z)\} - {11 \over 4} \alpha_S^{-1}
(M_Z) \right] < 3.81.
\end{equation}
Hence $\rm M_R < 4.1$ TeV and $\rm M_U > 1.1 \times 10^{16}$ GeV.  If the
central values of all three experimental inputs are used in the above,
$\rm M_R = 2.56$ TeV and $\rm M_U = 3.36 \times 10^{16}$ GeV would be obtained.
However, as in the SO(10) case, two-loop effects are significant and solving
Eq. (25) with the appropriate $b_{ij}$'s, it is found that
\begin{equation}
\rm M_R = 5.2~TeV, ~~~~~ M_U = 2.0 \times 10^{16}~GeV.
\end{equation}
This example is shown in Fig. 2.

A discrete $\rm Z_2$ symmetry is also assumed in this model.  The {\bf 16*}
and the $\Phi_{12} \sim (1,2,2,0)$ supermultiplets are even, and all the
others are odd.  The $\rm SU(2)_R \times U(1)_X$ gauge symmetry breaks down
to $\rm U(1)_Y$ through the nonzero vacuum expectation value of the
$\Phi_{\rm R}^c \sim (1,1,2,-1/2)$ component of the {\bf 16*}.
The breaking of $\rm SU(2)_L \times U(1)_Y$ down to $\rm U(1)_Q$
comes from $\Phi_{12}$ as well as $\Phi_{\rm L}^c \sim (1,2,1,1/2)$,
but the vacuum expectation value of the latter, which contributes to
the mass term $\nu N$, is assumed to be negligibly small in order to
have very small Majorana masses for the known neutrinos.  As discussed
earlier, this term was absent in the SO(10) model because it was
possible there to make $\Phi_{\rm L}^c$ odd.

The quarks and leptons within the three {\bf 16}'s or the {\bf 16*}
acquire masses through their couplings with the bidoublets, but the
{\bf 16}'s do not mix with the {\bf 16*} because of the assumed discrete
$\rm Z_2$ symmetry.  However, they do interact with the three singlet
$N$'s.  Hence an exotic $q'$ in the {\bf 16*} will decay into its
corresponding $q$ in the {\bf 16} and a virtual scalar $N$ which then
turns into a neutrino and a neutralino which has a $\Phi_{\rm L}^c$
component or a charged lepton and a chargino if kinematically allowed.
This differs from the usual models of mirror fermions,\cite{12}
{\it i.e.} fermions belonging to the {\bf 16*}, where they are
routinely assumed to mix with the ordinary ones.

In conclusion, it has been shown that it is possible to have an
$\rm SU(3) \times SU(2)_L \times SU(2)_R \times U(1)$ gauge symmetry
with $g_{\rm L} = g_{\rm R}$ at the TeV energy scale in at least two
supersymmetric models of grand unification, based on SO(10) and
$\rm SO(10) \times SO(10)$ respectively.  New particles are of course
necessary but there are simple solutions even though the possibility
of unification is very sensitive to small changes in the $b_i$'s.
See for example Eqs. (18) to (20).  The contributions of the new
particles come in large increments, so it is not possible to fine-tune
the $b_i$'s to get whatever values of $\rm M_U$ and $\rm M_R$ that may
be desired.  However, as it turns out in both models, the experimental
inputs of $\alpha^{-1}$, $\rm sin^2 \theta_W$, and $\rm \alpha_S$ at
$\rm M_Z$ do happen to require a few TeV for $\rm M_R$ and about
$10^{16}$ GeV for $\rm M_U$.  The models are also phenomenologically
natural and realistic.  Of particular note is the mass matrix of Eq. (29)
which results in the highly suppressed Majorana neutrino mass of Eq. (30).
New physics at the TeV energy scale beyond the minimal supersymmetric
standard model is clearly possible even in the face of grand unification.
\newpage
\begin{center} {ACKNOWLEDGEMENT}
\end{center}

I thank T. Yanagida for hospitality at Tohoku University during a recent
visit where I had many fruitful discussions with him and other colleagues
on the substance of this paper.  This work was supported in part by the
U. S. Department of Energy under contract No. DE-AT03-87ER40327.
\vspace{0.5in}

\bibliographystyle{unsrt}

\vspace{0.5in}

\begin{center} {FIGURE CAPTIONS}
\end{center}

\noindent Fig. 1.  Evolution of $\alpha_i^{-1}$ in the SO(10) model with
$\rm M_R = 1~TeV$ and $\rm M_U = 1.0 \times 10^{16}~GeV$.

\noindent Fig. 2.  Evolution of $\alpha_i^{-1}$ in the $\rm SO(10) \times
SO(10)$ model with $\rm M_R = 5.2~TeV$ and $\rm M_U = 2.0 \times 10^{16}$
GeV.

\end{document}